\documentclass[twocolumn,showpacs,preprintnumbers,amsmath,amssymb]{revtex4}

\usepackage{graphicx}
\usepackage{epsfig}
\usepackage{rotating}
\usepackage{dcolumn}
\usepackage{bm}

\begin{document}

\preprint{HEP/123-qed}
\title{Beam-Helicity Asymmetries in Double Pion Photoproduction off the Proton}
\author{
  D.~Krambrich$^1$,
  F.~Zehr$^2$,
  A. Fix$^{18}$,
  L. Roca$^{19}$,
  P.~Aguar$^{1}$,
  J.~Ahrens$^{1}$,
  J.R.M.~Annand$^{3}$,
  H.J.~Arends$^{1}$,
  R.~Beck$^{1,4}$,
  V.~Bekrenev$^{5}$,
  B.~Boillat$^{2}$,
  A.~Braghieri$^{6}$,
  D.~Branford$^{7}$,
  W.J.~Briscoe$^{8}$,
  J.~Brudvik$^{9}$,
  S.~Cherepnya$^{10}$,
  R.~Codling$^{3}$,
  E.J.~Downie$^{3}$,
  P.~Dexler$^{11}$,
  D.I.~Glazier$^{7}$,
  P.~Grabmayr$^{12}$,
  R.~Gregor$^{11}$,
  E.~Heid$^{1}$,
  D.~Hornidge$^{13}$,
  O.~Jahn$^{1}$,
  V.L.~Kashevarov$^{10}$,
  A.~Knezevic$^{14}$,
  R.~Kondratiev$^{15}$,
  M.~Korolija$^{14}$,
  M.~Kotulla$^{2}$,
  B.~Krusche$^{2}$,
  A.~Kulbardis$^{5}$,
  M.~Lang$^{1,4}$,
  V.~Lisin$^{15}$,
  K.~Livingston$^{3}$,
  S.~Lugert$^{11}$,
  I.J.D.~MacGregor$^{3}$,
  D.M.~Manley$^{16}$,
  M.~Martinez$^{1}$,
  J.C.~McGeorge$^{3}$,
  D.~Mekterovic$^{14}$,
  V.~Metag$^{11}$,
  B.M.K.~Nefkens$^{9}$,
  A.~Nikolaev$^{1,4}$,
  P.~Pedroni$^{6}$,
  F.~Pheron$^{2}$,
  A.~Polonski$^{15}$,
  S.N.~Prakhov$^{9}$,
  J.W.~Price$^{9}$,
  G.~Rosner$^{3}$,
  M.~Rost$^{1}$,
  T.~Rostomyan$^{6}$,
  S.~Schumann$^{1,4}$,
  D.~Sober$^{17}$,
  A.~Starostin$^{9}$,
  I.~Supek$^{14}$,
  C.M.~Tarbert$^{7}$,
  A.~Thomas$^{1}$,
  M.~Unverzagt$^{1,4}$,
  Th.~Walcher$^{1}$,
  D.P.~Watts$^{7}$\\
(The Crystal Ball at MAMI, TAPS, and A2 Collaborations)
}
\affiliation{
  $^{1}$\mbox{Institut f\"ur Kernphysik, University of Mainz, Germany}\\
  $^{2}$\mbox{Department Physik, Universit\"at Basel, Switzerland}\\
  $^{3}$\mbox{Department of Physics and Astronomy, University of Glasgow, Glasgow, UK}\\
  $^{4}$\mbox{Helmholtz-Institut f\"ur Strahlen- und Kernphysik, University Bonn, Germany}\\
  $^{5}$\mbox{Petersburg Nuclear Physics Institute, Gatchina, Russia}\\
  $^{6}$\mbox{INFN Sezione di Pavia, Pavia, Italy}\\
  $^{7}$\mbox{School of Physics, University of Edinburgh, Edinburgh, UK}\\
  $^{8}$\mbox{Center for Nuclear Studies, The George Washington University, Washington, DC, USA}\\
  $^{9}$\mbox{University of California at Los Angeles, Los Angeles, CA, USA}\\
  $^{10}$\mbox{Lebedev Physical Institute, Moscow, Russia}\\
  $^{11}$\mbox{II. Physikalisches Institut, University of Giessen, Germany}\\
  $^{12}$\mbox{Physikalisches Institut Universit\"at T\"ubingen, T\"ubingen, Germany}\\
  $^{13}$\mbox{Mount Allison University, Sackville, NB, Canada}\\
  $^{14}$\mbox{Rudjer Boskovic Institute, Zagreb, Croatia}\\
  $^{15}$\mbox{Institute for Nuclear Research, Moscow, Russia}\\
  $^{16}$\mbox{Kent State University, Kent, OH, USA}\\
  $^{17}$\mbox{The Catholic University of America, Washington, DC, USA}\\
  $^{18}$\mbox{Laboratory of Mathematical Physics, Tomsk Polytechnic University, 634034 Tomsk, Russia}\\ 
  $^{19}$\mbox{Departamento de Fisica, Universidad de Murcia, Murcia, Spain}
  }
\date{\today}

\begin{abstract}
Beam-helicity asymmetries have been measured at the MAMI accelerator in Mainz 
in the three isospin channels 
$\vec{\gamma}p\rightarrow \pi^{+}\pi^0n$, 
$\vec{\gamma}p\rightarrow \pi^{0}\pi^0p$ and 
$\vec{\gamma}p\rightarrow \pi^{+}\pi^{-}p$ . 
The circularly polarized photons, produced from bremsstrahlung 
of longitudinally polarized electrons, were tagged with the Glasgow magnetic 
spectrometer. Charged pions and the decay photons of  $\pi^0$ mesons were 
detected in a $~4\pi$ electromagnetic calorimeter which combined the Crystal 
Ball detector with the TAPS detector. The precisely measured asymmetries are 
very sensitive to details of the production processes and are thus key 
observables in the modeling of the reaction dynamics.
\end{abstract}
\pacs{PACS numbers: 
13.60.Le, 14.20.Gk, 14.40.Aq, 25.20.Lj
}

\maketitle

Double pion photoproduction allows the study of sequential decays of nucleon 
resonances via intermediate excited states, as well as the coupling of nucleon 
resonances to $N\rho$ and $N\sigma$. It has therefore become an attractive tool
for the study of the excitation spectrum of the nucleon, which is intimately
connected to the properties of QCD in the non-perturbative range. 
Its contribution to the total electromagnetic response of the nucleon is 
substantial.
In the second resonance region, comprising the $P_{11}$(1440), $S_{11}$(1535),
and $D_{13}$(1520) resonances, roughly 50\% of the total photoabsorption 
cross section originates from it. In this energy region total cross sections and 
invariant mass distributions of the $\pi\pi$- and the $\pi N$-pairs have been
measured with the DAPHNE and TAPS detectors at the MAMI accelerator in Mainz 
\cite{Braghieri_95,Haerter_97,Zabrodin_97,Zabrodin_99,Wolf_00,Kleber_00,Langgaertner_01,Kotulla_04},
at GRAAL in Grenoble (also linearly polarized beam asymmetry) \cite{Assafiri_03,Ajaka_07}, with the CLAS detector 
at JLab (electroproduction) \cite{Ripani_03}, and at ELSA in Bonn 
\cite{Thoma_07,Sarantsev_07}. 
More recently, also polarization observables have been measured at the MAMI 
accelerator \cite{Ahrens_03,Ahrens_05,Ahrens_07} and at the CLAS facility
at Jlab \cite{Strauch_05}.

In spite of all these efforts, even the interpretation of the data in the
energy region, where only few resonances can contribute, is still 
surprisingly controversial \cite{Krusche_03} since the available data
do not sufficiently constrain the model analyses. It is thus evident that 
the search for missing resonances at higher energy requires a better 
understanding of the reaction mechanisms.
The controversy has far reaching consequences not only for the 
N$^{\star}$ excitation spectrum itself, but as discussed below also in the 
field of the much discussed hadron in-medium 
properties. 

There is agreement that the $\pi^+\pi^-$ final state is 
dominated by background terms in particular of the $\Delta$-Kroll-Rudermann 
type, while $\pi^0\pi^0$ has only small background contributions and thus is
particularly suited for the study of sequential resonance decays.
However, even for the latter the results of different reaction models are
contradictory. Calculations by the Valencia group 
\cite{Gomez_96,Nacher_01,Nacher_02} emphasize a large contribution from the 
$D_{13}\to \Delta\pi^0\to p\pi^0\pi^0$ decay. Laget and coworkers 
\cite{Assafiri_03}, instead find a dominant contribution from the 
$P_{11}$(1440)$\rightarrow N\sigma$ decay 
and a recent analysis by the Bonn-Gatchina group \cite{Thoma_07,Sarantsev_07} 
reports a strong contribution from the $D_{33}$(1700) resonance, which is not
seen in other models. Modifications of the invariant mass distributions
of the $\pi^0$ pairs for photoproduction off heavy nuclei have been discussed
in view of the predicted $\sigma$ in-medium modification resulting from
partial chiral symmetry restoration \cite{Messchendorp_02,Bloch_07}, 
however, a better understanding of the elementary production 
processes is obligatory. Similarly, for the mixed charge channel $n\pi^+\pi^0$
all early model calculations (see e.g. \cite{Gomez_96}) failed already
in the reproduction of the total cross section. Only the introduction of
a strong contribution from the $\rho$ meson \cite{Nacher_01,Nacher_02,Fix_05}, 
motivated by the shape of the measured invariant mass distributions 
\cite{Zabrodin_99,Langgaertner_01}, improved the situation. Again, a close 
connection to a different problem, namely the still unexplained strong
suppression of the second resonance bump in photoproduction off nuclei
(see e.g. \cite{Bianchi_94}) is involved, where a possible explanation might
arise from the in-medium modification of the $D_{13}(1520)\rightarrow N\rho$
decay \cite{Langgaertner_01}.  

\begin{figure}[th]
\begin{center}
\epsfig{figure=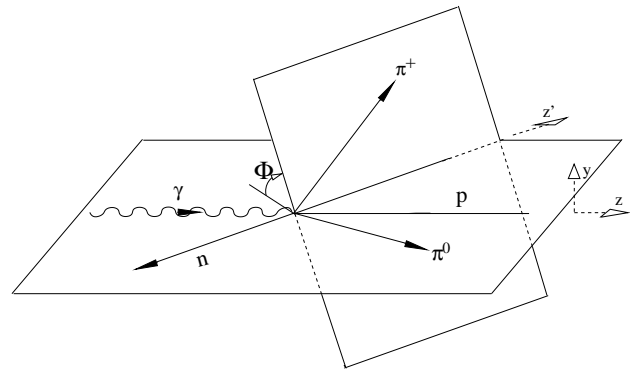,width=7.5cm}  
\caption{Vector and angle definitions. $\Phi$ is the angle 
between the reaction plane (defined by $\vec{k}$ and $\vec{p}_{n}$) and the 
production plane of the two pions (defined by
$\vec{p}_{\pi^0}$ and $\vec{p}_{\pi^+}$). 
\label{fig:asym} 
}
\end{center}
\end{figure}

Recently, model predictions
\cite{Nacher_02,Roberts_05,Fix_05,Roca_05}, which indicated that polarization
observables are extremely sensitive for the disentanglement of the  
reaction mechanisms, have triggered wide-spread experimental activities. 
The advent of accelerators with highly polarized electron beams has provided a new 
tool for this field: meson photoproduction using circularly polarized photons. They 
are produced by the bremsstrahlung of longitudinally polarized electrons in an 
amorphous radiator. The polarization transfer obeys a simple formula given by Olsen 
and Maximon \cite{Olsen_59}. The beam helicity asymmetry can then be measured by 
comparing the event rates for the two helicity states of the beam.
Parity conservation precludes any sensitivity of the cross section in a
two-body reaction to beam helicity alone, but in a reaction with three or more 
particles in the final state, circularly polarized
photons can lead to asymmetries even for an unpolarized target.
Until recently there was little effort to study these effects until two
experimental programs at JLab observed strong signals. In hyperon
photoproduction, the decay of the final state $\Lambda$ or $\Sigma$ hyperon has an
angular dependence on the hyperon polarization, and a recent experiment
\cite{Bradford_07} has shown that the polarization transfer along the photon momentum
axis is nearly 100\%. In an analysis of charged double-pion production 
$\gamma p\to p\pi^+\pi^-$ measured with CLAS, Strauch {\it et al.} 
\cite{Strauch_05} found a large helicity asymmetry in the distribution of 
$\Phi$, the angle between the two-pion plane and the $\gamma p$ reaction plane
(see Fig.~\ref{fig:asym}). 

The Crystal-Ball-TAPS collaboration at the Mainz microtron MAMI \cite{Walcher_90}
has recently taken data on the photoproduction of the three $\pi\pi N$ final states 
accessible with a proton target: $\gamma p\to p\pi^+\pi^-$, $\gamma p\to p\pi^0\pi^0$, 
and $\gamma p\to n\pi^+\pi^0$, using circularly polarized photons. This Letter presents 
the beam-helicity asymmetries in a form similar to that of Strauch {\it et al.} 
The data were taken with tagged photons incident on a 4.8 cm 
long liquid hydrogen target (surface density 0.201 nuclei/barn). Contributions 
from the target windows (2$\times$60 $\mu$m Kapton) were determined with empty 
target measurements and subtracted. 
The photons of up to 820 MeV, were produced by the bremsstrahlung of 883 MeV
longitudinally polarized electrons. The energy of the photons was determined 
by the Glasgow photon tagger \cite{Anthony_91,Hall_96} with a resolution of 
approximately 2 MeV full width. The target was located inside the Crystal Ball 
(CB) \cite{Starostin_01}, consisting of 672 NaI crystals that 
covered the full azimuthal range for polar angles between 20$^{\circ}$ - 160$^{\circ}$. 
The angular region from 20$^{\circ}$ down to
1$^{\circ}$ was covered by the TAPS detector \cite{Novotny_91,Gabler_94} with 
510 BaF$_2$ crystals arranged as a hexagonal wall. The target was surrounded by a 
{\bf P}article {\bf I}dentification {\bf D}etector (PID) \cite{Watts_04} and two cylindrical 
multiple wire proportional chambers (MWPC) \cite{Audit_91}. Protons and charged pions hitting 
the CB were identified by an $E-\Delta E$ analysis, using the energy information of the CB and 
the PID. 
For TAPS, the separation of photons, neutrons, protons, and charged pions can be
achieved in principal as discussed in \cite{Bloch_07}. However, here these methods were 
only used for a clean identification of photons. Protons and charged pions in TAPS
were not included in the analysis, since 
their separation was less clean than the $E-\Delta E$ analysis 
by PID and CB.

\begin{figure}[thb]
\begin{center}
\epsfig{figure=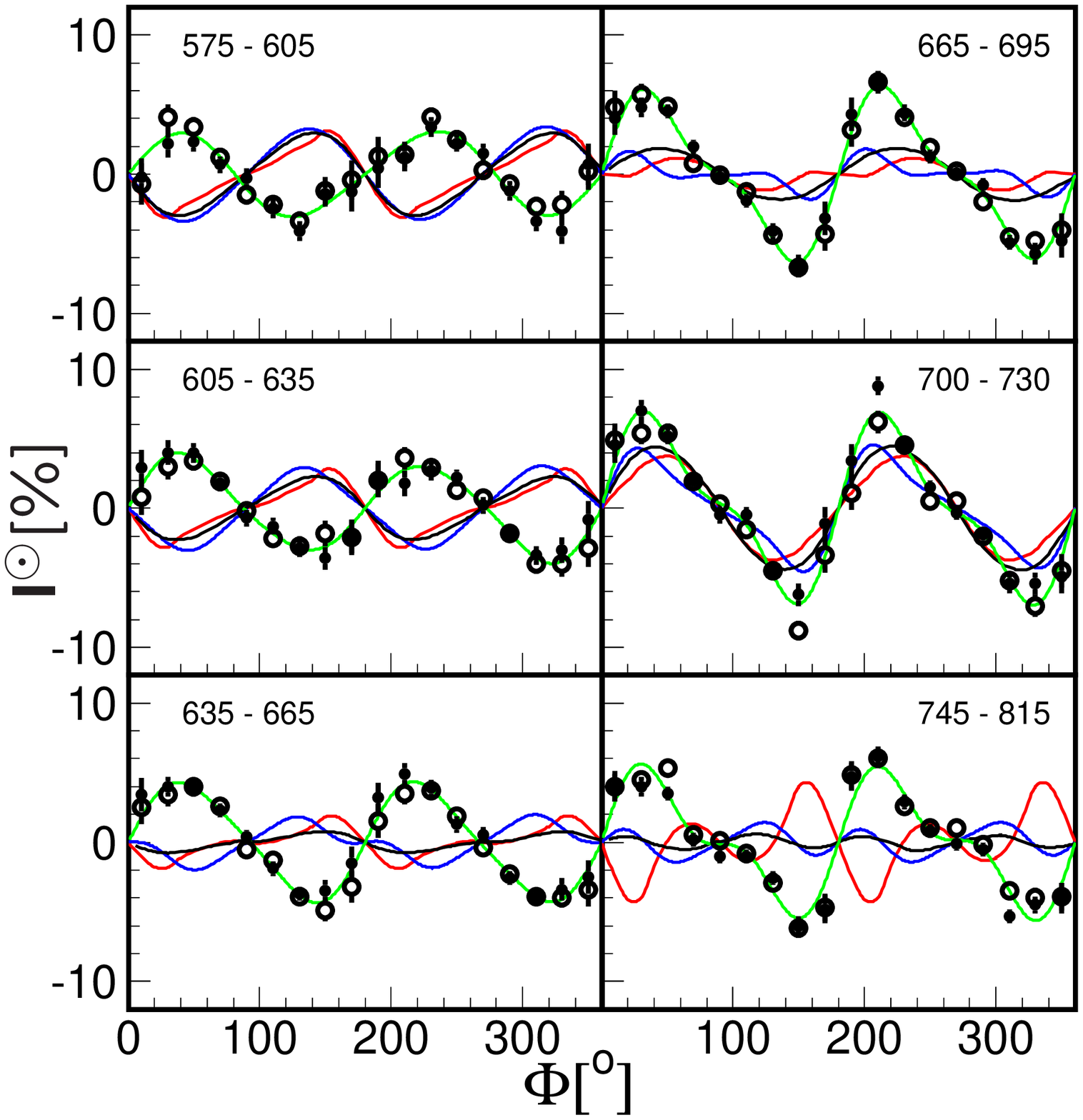,width=8.1cm}\\
\epsfig{figure=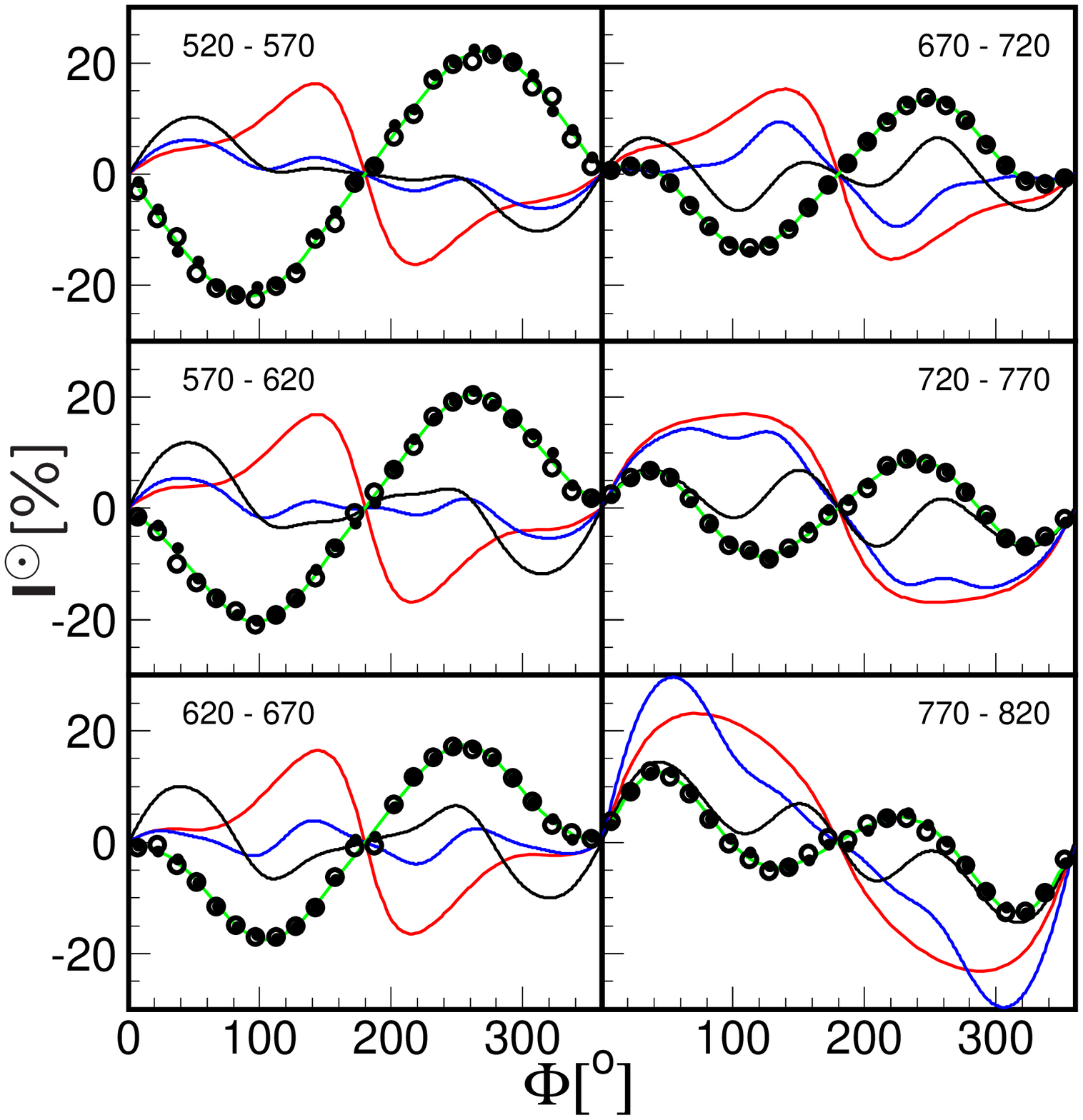,width=8.1cm}
\caption{upper part:  Beam-helicity asymmetry in the 
$\vec{\gamma}p\rightarrow p\pi^{+}\pi^-$ reaction for different bins of photon 
energy.
Filled symbols: $I^{\odot}(\Phi)$, open symbols: 
$-I^{\odot}(2\pi-\Phi)$. Green curves: fit to the data. 
Red curves: Fix and Arenh\"ovel model
\cite{Fix_05}; Blue:  Roca \cite{Roca_05}, Black: Roca \cite{Roca_05}
for 4$\pi$ acceptance. Bottom part: Beam-helicity 
asymmetry for $\vec{\gamma}p\rightarrow n\pi^{+}\pi^{0}$. Notation as for 
left-hand side except black curves: Roca \cite{Roca_05} without 
$D_{13}\to N\rho$.
\label{fig:pi_char}
}
\end{center}
\end{figure} 

In the first step of the reaction identification for the $p\pi^0\pi^0$ final 
state, events with four photons and one or no proton candidate were selected. 
Similarly, for the $n\pi^+\pi^0$ final state two photons, a $\pi^+$, and one or 
no neutron were required. The $\pi^0$ mesons where then identified
by a standard invariant mass analysis.
Further identification of the reactions was based on missing mass 
analyses for the recoil nucleons in a manner similar to that described in
\cite{Wolf_00,Kleber_00,Langgaertner_01,Kotulla_04}.
It was used to remove small residual background from $\eta\to 3\pi$ decays, 
which however, was much less important than in previous experiments, since due 
to the large solid angle coverage, in most cases the third pion was also seen.
For both reactions, the recoil nucleons were treated as missing particles, 
no matter whether a candidate was found or not, so that the results are 
independent of the detector acceptance and efficiency for recoil nucleons.
This event selection guaranteed full solid angle coverage for the $\pi^0\pi^0$
channel. For the $\pi^+\pi^0$  reaction only events with the $\pi^+$ at laboratory 
polar angles smaller than 20$^{\circ}$  (and larger than 160$^{\circ}$) were excluded, 
which has a negligible effect on the measured asymmetries. 
Since identification of the double charged channel is missing the powerful
tool of invariant mass analysis, in this case detection of all three
charged particles was required in order to achieve an equally good
background suppression as for the other channels (residual background from
$\gamma p\rightarrow p\pi^+\pi^-\pi^0$ was again removed with a missing mass analysis). 
This selection means that for the double charged channel only events 
with all three particles at laboratory polar angles larger than 20$^{\circ}$ 
(and smaller than 160$^{\circ}$) were accepted. This limitation was accounted for
in the model calculations. 

\begin{figure}[ttt]
\begin{center}
\epsfig{figure=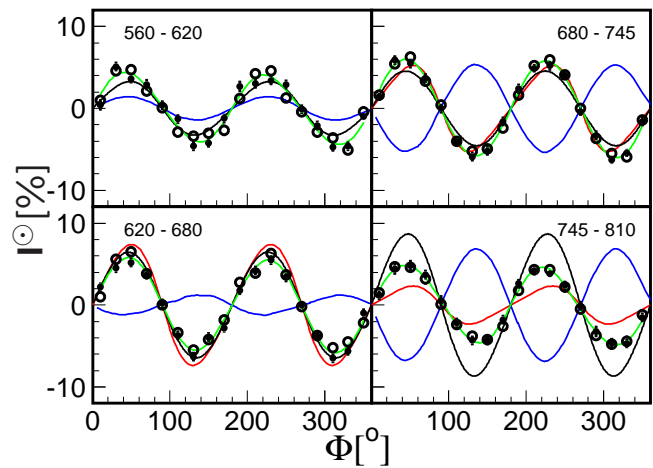,width=8.5cm} 
\caption{Beam-helicity asymmetry for 
$\vec{\gamma}p\rightarrow \pi^{0}\pi^{0}p$. Notation as 
Fig.~\ref{fig:pi_char} except black curves: Bonn-Gatchina model 
\cite{Thoma_07,Sarantsev_07} 
\label{fig:pi_neu}
}
\end{center}
\end{figure} 

The missing mass spectra for all three reactions were
extremely clean and very well reproduced by Monte Carlo simulations.  
Residual background was estimated at maximum at the few per cent level 
(certainly well below 5 \% for all channels) and is thus not relevant for any 
results presented here. Details of the analysis will be discussed in an upcoming 
paper about total cross sections and invariant mass distributions. 

In a reaction produced by circularly polarized photons on an unpolarized target
the beam-helicity asymmetry $I^{\odot}$ is defined by:
\begin{equation}
I^{\odot}(\Phi)=\frac{1}{P_{\gamma}}
	        \frac{d\sigma^{+}-d\sigma^{-}}{d\sigma^{+}+d\sigma^{-}}
	       =\frac{1}{P_{\gamma}}
                \frac{N^{+}-N^{-}}{N^{+}+N^{-}}
\label{eq:circ}		
\end{equation}
where $d\sigma^{\pm}$ is the differential cross section for each of the two
photon helicity states, and $P_{\gamma}$ is the degree of circular polarization 
of the photons. The latter is calculated as product of the polarization degree of
the longitudinally polarized electrons (82$\pm$5)\% and the photon-energy-dependent 
polarization transfer factor \cite{Olsen_59}. 
In the energy range of interest, $P_{\gamma}$
was between 60\% and 80\%. Possible differences in the number of 
incident photons for the two helicity states have been determined to be at the 
5$\times$10$^{-4}$ level, i.e. they are negligible. The angle $\Phi$ between
reaction and production plane is calculated as defined in the work of Roca 
\cite{Roca_05} from the three-momenta of the particles (the same construction 
was used for the analysis of the CLAS-data \cite{Strauch_05}). 
For $\pi^+\pi^0$ production the two pions are ordered as shown in Fig.~\ref{fig:asym}.
For double $\pi^0$ production and
for the double charged state their assignment is randomized since 
the experiment cannot distinguish positively and negatively charged pions.
This means that for the latter two $I^{\odot}(\Phi)=I^{\odot}(\Phi +\pi)$. 

For the extraction of the asymmetry
$I^{\odot}(\Phi,\Theta_{\pi_1},\Theta_{\pi_2},...)$ in a limited region of 
kinematics, the differential cross sections $d\sigma^{\pm}$ can be  
replaced by the respective count rates $N^{\pm}$  (right hand side of
Eq.~\ref{eq:circ}), since all normalization factors cancel in the ratio. 
In principle efficiency weighted count rates ought to be used to obtain the 
angle integrated asymmetries. For the two final states $\pi^0\pi^0$ and 
$\pi^+\pi^0$, for which also precise total cross sections and invariant mass 
distributions will be published elsewhere, the detection efficiency was 
modeled with Monte Carlo simulations. However, since the efficiencies are 
rather flat functions of the pion polar angles, the effect of the efficiency
corrections on the asymmetries was negligible. As in the CLAS 
experiment \cite{Strauch_05} only the raw asymmetries are given for $\pi^+\pi^-$ 
production. 

The measured asymmetries are summarized in 
Figs.~\ref{fig:pi_char}, \ref{fig:pi_neu} as functions of $\Phi$.  
Parity conservation enforces 
$I^{\odot}(\Phi)=-I^{\odot}(2\pi-\Phi)$. This condition was not used as a 
constraint in the analysis but is very well respected, demonstrating the 
excellent quality of the data. 

The asymmetries are compared to the results from the model of  
Fix and Arenh\"ovel \cite{Fix_05} and the Valencia model 
\cite{Roca_05}, which were calculated taking into account 
the acceptance limitations for $\pi^+\pi^-$ and the fact that 
$\pi^-$ could not be distinguished from $\pi^+$ in the detectors. 
For this channel also the prediction of the Valencia model
for full 4$\pi$ acceptance is shown. At least in the framework of the model,
the effect from the acceptance limitation is small. 
A similar result as in the CLAS experiment
\cite{Strauch_05} is found. The two models make similar predictions, but agree
with the measurements only in the energy range around 715 MeV. For 
$n\pi^0\pi^+$, the model results are similar above 700 MeV, but are nowhere 
in agreement with the data. For the Valencia model \cite{Roca_05} also the 
solution without the $D_{13}\to N\rho$ contribution is shown. It was introduced
into the model \cite{Nacher_01,Nacher_02} in order to reproduce the previously
non-understood total cross section and pion invariant mass distributions 
\cite{Zabrodin_99,Langgaertner_01}. However, in the $D_{13}$ range, inclusion
of this contribution does not at all improve the agreement with the asymmetries.
Finally, for $p\pi^0\pi^0$ Fix's model and the Bonn-Gatchina analysis
(not available for the other iso-spin channels) \cite{Thoma_07,Sarantsev_07} are 
in fairly good agreement with the data, while the Valencia model is out of phase.  

\begin{figure}[ttt]
\begin{center}
\epsfig{figure=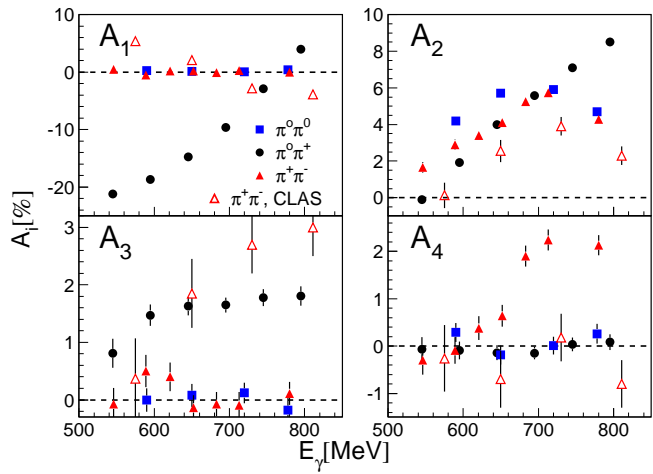,width=8.5cm} 
\caption{Fitted coefficients of the expansion of the beam-helicity asymmetries.
Note that the acceptance for the present
$\pi^+\pi^-$ and the CLAS data are different (see text).  
\label{fig:coeff}
}
\end{center}
\end{figure} 

Due to its symmetry $I^{\odot}(\Phi)$ can be expanded into a sine-series
(odd coefficients vanish for identical pions):
\begin{equation}
I^{\odot}(\Phi)=\sum_{n=1}^{\infty}A_{n}\mbox{sin}(n\Phi)
\label{eq:series}
\end{equation}
The data have have been fitted to Eq.~\ref{eq:series} for $n\leq 4$ 
(higher orders were not significant), and the results are summarized in 
Fig.~\ref{fig:coeff}. 
For the $p\pi^0\pi^0$ and $p\pi^+\pi^-$ final states the results for the odd 
terms $A_1$ and $A_3$ are consistent with zero, which is additional evidence 
that no false asymmetries have been generated in the experiment. 
For comparison the CLAS results \cite{Strauch_05} for $p\pi^+\pi^-$
have also been fitted. Since in the CLAS experiment negatively and positively 
charged pions were distinguished, the odd terms can also contribute, but the 
even terms $A_2$ and $A_4$ can be compared to the present results. 
One must, however, keep in mind that the acceptance was not 
identical (CLAS covered polar angles down to 8$^{\circ}$, this experiment 
down to 20$^{\circ}$). For $A_2$ the energy dependence is similar, although 
the present values are somewhat larger. No significant contribution from $n=4$ 
was found for the CLAS experiment, but in the present measurement it contributes
up to 2\%. The comparison of the results
for the three final states highlights the different reaction mechanisms in the 
three isospin channels.  

In summary, precise measurements of the beam-helicity asymmetry for double pion 
photoproduction on the proton have been presented for all three isospin channels. 
The comparison with model predictions highlights, both the challenges and potential 
rewards for the extraction of resonance properties. On the one hand, the progress 
in experimental techniques allows precise measurements of this observable, 
and the predictions for it are very sensitive to the internal mechanisms of 
the models. On the other hand, 
the general lack of agreement between experiment and theory signals 
that significant improvements in the models are needed. 
The present data can provide rigorous tests for future developments on 
the way to an eventual reliable extraction of resonance contributions
from double pion photoproduction. The very precise results for the total cross sections,
including sensitive measurements of the threshold behavior in view of the predictions of
chiral perturbation theory, and the invariant mass distributions of pion - pion and
pion - nucleon pairs, which have been extracted with a precision far superior to any
previous measurements, will be presented in an upcoming paper.

\acknowledgments
We wish to acknowledge the outstanding support of the accelerator group 
and operators of MAMI. This work was supported by Deutsche
Forschungsgemeinschaft (SFB 443, SFB/TR 16), DFG-RFBR (Grant No. 05-02-04014),
Schweizerischer Nationalfonds, UK Engineering and Physical Sciences Research 
Council, European Community-Research Infrastructure 
Activity (FP6), the US DOE, US NSF and NSERC (Canada).
We thank the undergraduate students of Mount Allison and George Washington  
Universities for their assistence.

\end{document}